\documentclass[12pt]{iopart}
\usepackage{graphicx}
\begin{document}

\title{Oscillations in molecular motor assemblies}

\author{Andrej Vilfan$^\dag$ and Erwin Frey$^\ddag$}

\address{\dag J. Stefan Institute, Jamova 39, 1000 Ljubljana,
  Slovenia}

\address{\ddag Arnold Sommerfeld Center for Theoretical Physics and
  Center of Nanoscience, Department of Physics,
  Ludwig-Maximilians-Universit\"at M\"unchen, Theresienstrasse 37,
  D-80333 M\"unchen, Germany}

\renewcommand{\Re}{{\rm Re\,}}
\renewcommand{\Im}{{\rm Im\,}}

%\ead{andrej.vilfan@ijs.si}

\begin{abstract}
  Autonomous oscillations in biological systems may have a biochemical
  origin or result from an interplay between force-generating and
  visco-elastic elements. In molecular motor assemblies the
  force-generating elements are molecular engines and the
  visco-elastic elements are stiff cytoskeletal polymers. The physical
  mechanism leading to oscillations depends on the particular
  architecture of the assembly. Existing models can be grouped into
  two distinct categories: systems with a {\em delayed force
    activation} and {\em anomalous force-velocity relations}.  We
  discuss these systems within phase plane analysis known from the
  theory of dynamic systems and by adopting methods from control
  theory, the Nyquist stability criterion.
\end{abstract}

\submitto{\JPCM} 

\maketitle

\section{Introduction}

Oscillations are ubiquitous in biological systems. Examples range from
circadian rhythms associated with external periodicities to autonomous
bio-chemical~\cite{Murray-book1} and mechano-chemical oscillators. Voltage
oscillations in systems of ion channels embedded in a cell membrane are one of
the most prominent examples for a biochemical oscillator. At the heart of its
explanation within the Hodgkin-Huxley model~\cite{Hodgkin.Huxley1952} there is
a negative differential current-voltage relation for the ion channels. Such
biochemical oscillators are by now well studied \cite{Murray-book1}. Recently,
there has been growing interest in mechano-chemical systems composed of
visco-elastic biopolymer arrays and active molecular machines driven by the
chemical cycle of ATP hydrolysis; these will be the focus of our contribution.
We ask how the interplay between force-generation and energy-input by the
molecular motors and the restoring forces and damping mechanisms provided by
the biopolymer system may lead to oscillatory behavior.

There are many examples of functional units in cellular systems whose main
components are molecular machines and visco-elastic elements. The part list of
{\em auditory hair bundles} contains stereocilia (elastic rods composed of
bundles of stiff biopolymers F-actin), myosin (a force-generating molecular
motor) and mechanosensitive ion-channels.  Spontaneous oscillations in such
systems are by now well documented experimentally \cite{Martin.Hudspeth2003}
and explained within simple theoretical models
\cite{Choe.Hudspeth1998,vilfan2003a}. This oscillatory behavior of hair bundles
provides a mechanism for active amplification of acoustic signals
\cite{Camalet.Prost2000}.  Another example are {\em eucaryotic cilia and
  flagella} which are used by many small organisms to swim. Here the main
structural element are axonemes, which consist of a cylindrical arrangement of
elastically linked microtubules (another example for a stiff biopolymer) and an
assembly of dynein motors located between neighboring microtubules. Forces
generated by the dynein motors induce relative sliding between microtubules
which in turn results into bending of the axonemes \cite{Brokaw1971}.
Oscillatory behavior in such systems is generic and intimately connected with a
Hopf bifurcation \cite{camalet99}. {\em Sarcomeres}, the contractile units of
muscle, are composed of two sets of protein filaments: myosin filaments
(containing myosin motors) and actin filaments. The shortening of the sarcomere
is achieved by the actin and myosin filaments sliding over one another.  While
the primary function of most types of muscle is to generate a unidirectional
force, some muscle (notably the asynchronous, also known as fibrillar or
myogenic insect flight muscle) are specialized on autonomous oscillatory
contraction.  The oscillation mechanism usually includes a number of regulatory
proteins that constitute the delayed stretch-activation mechanism.  However,
oscillations were also observed in skeletal muscle fibrils, mostly under
non-physiological conditions \cite{Okamura.Ishiwata1988}.  Theoretical studies
show that even skeletal muscle myosin could produce oscillatory motion
\cite{duke99,duke2000,vilfan2003b}, which is normally not observable because
the contributions of different sarcomeres cancel out.  However, a sudden change
in applied load can lead to synchronization of sarcomeres and the oscillations
become observable for a certain period of time \cite{Edman.Curtin2001}.
Assemblies of molecular motors and stiff biopolymers also play an essential
role during cell division. The duplicated genome is segregated into two
daughter cells through the action of the mitotic spindle. It consists of fibres
(microtubule bundles) radiating from two poles and meeting at the equator in
the middle. The spindle poles have been found to oscillate
\cite{Grill.Hyman2001}. Recently it has been shown that these oscillations are
again due to an interplay between force-generating and elastic elements
\cite{Grill.Julicher2005}.

Here we will focus on a discussion of the generic features of oscillations in
molecular motor assemblies to highlight the common physical mechanisms. For
more system specific and detailed discussions we refer the reader to the
literature. The defining elements of molecular motor assemblies are: (i) an
external energy source driving the system towards a non-equilibrium steady
state, (ii) viscoelastic elements providing restoring forces and damping
mechanisms, (iii) possibly force-dependent biochemical reactions which allow
for switching between different states of the molecular motor assembly.  The
particular interconnection between these components gives rise to the system
response which may be controlled by the biochemical reaction rates. Defined as
such there are obvious analogies to system and control theory which is a well
established discipline in engineering \cite{Dorf-book}. In particular, linear
system analysis is a most fruitful concepts which we will also employ to
characterize the system's dynamic response.

The mechanisms for oscillations in chemochemical models discussed in the
literature can be grouped into two distinct categories. In {\em delayed force
  activation} systems an imposed displacement $x$ is transduced into a force
$f$ with some delay time $\tau$ (Debye relaxator). As such the system would
evolve towards a time-independent steady state. Oscillatory response is
achieved only after coupling the Debye relaxator to other dynamic elements with
either a negative stiffness or some inertial (massive) load.  {\em Anomalous
  force-velocity relations} are an alternative route towards oscillatory
behavior.  This means that an assembly of motors can move with two different
velocities (e.g., one positive and one negative) under the same load
\cite{riveline98}.  If attached to an external spring, the motors will move
forwards until the spring force exceeds a certain threshold, then they will
switch to the other stable state and slide backwards, until another threshold
is reached and so the cycle repeats \cite{juelicher97}.

\section{Response Functions}

To assess the stability of an active mechanical system like an insect muscle
attached to the wing or the dynein motor acting between two tubulin filaments
in a flagellum, the response function can be defined the following way.  We
connect the active system to a mechanical actuator that keeps its length at the
desired point of operation and at the same time imposes small oscillations with
the amplitude $x_0$ around that point
\begin{equation}
  \label{eq:oscillating1}
  \Delta x(t) = \Re x_0 e^{ i \omega t}\;.
\end{equation}
Fur sufficiently small oscillations the system responds with a force
that oscillates with the same frequency
\begin{equation}
  \label{eq:oscillating2}
  \left< \Delta f (t) \right> = \Re f_0 e^{ i \omega t}\;.
\end{equation}
Here $\left< \right>$ denote an ensemble average over different
realizations of the response of the stochastic system. Note that the
sign in the exponent of the Fourier transform was chosen in the way
that is most common in the literature on muscle, although it is
opposite to the convention frequently used in physics. We define the
force in the way that it is positive if the motor system is being
pulled upon in the direction of positive $x$.  In the limit of small
amplitude $x_0$, the relation becomes linear
\begin{equation}
  \label{eq:lin_response}
  f_0=G(\omega) x_0 \;,
\end{equation}
with the frequency dependent response function or modulus $G(\omega)$.
Here the real part of $G$ corresponds to the elastic and the imaginary
part to the dissipative response. For a spring with a spring constant
$k$ the response function is $G(\omega)=k$ and for a damping element
(``dashpot'') $G(\omega)=i \omega \gamma$.

One may also encounter a situation where for a given force $f$ one is
observing the displacement $x$ for various realizations of the system
response,
\begin{equation}
  \label{eq:susceptibility}
  \left<x (\omega) \right> = \chi (\omega) \, f(\omega) \;.
\end{equation}
The two response functions become equivalent
\begin{equation}
  \chi(\omega)=1/G(\omega)
\end{equation}
if the following two conditions are fulfilled.  First, the system
needs to be stable, so that the operational point is the same
regardless whether the position or the force is imposed. Second, the
systems are large enough such that variance of the stochastic variable
is small compared to its mean.

\section{The Nyquist stability criterion}

The Nyquist stability criterion is a convenient tool to analyze the stability
of a dynamical system if the complex response function $G(\omega)$ is known
over the whole frequency range. It states that the system is dynamically
unstable if the plot of $\Im G(\omega)$ vs. $\Re G(\omega)$ (Nyquist plot,
sometimes also called Cole-Cole plot) encircles the coordinate origin in
clockwise direction.  It is frequently used in engineering, especially for the
stability analysis of closed feedback loops in electrical
circuits \cite{Dorf-book}.

To derive it we first note that the system is dynamically unstable if it starts
oscillating with a growing amplitude in the absence of a force, $\left<f
\right>=0$. This is the case if there exists an eigenmode, i.e., a
complex solution of the equation $\left< f (x_0,\omega) \right>=0$ with $\Im
\omega <0$, since then the corresponding solution $\Delta x(t)\propto
e^{i\omega t}$ obviously diverges for $t\to +\infty$.  This means that
$G(\omega)$ has a zero and $\chi(\omega)$ has a pole in the lower half-plane
with $\Im \omega < 0$.  The Nyquist criterion is based on the fact that
$G(\omega)$ is an analytical complex function and utilizes Cauchy's theorem to
relate the number of zeros in the negative imaginary $\omega$ half-plane to the
contour of $G(\omega)$, evaluated for real values of $\omega$.

According to Cauchy's theorem, the change of phase along a closed
contour in clockwise direction in the complex $\omega$ plane equals
\begin{equation}
  \label{eq:cauchy}
  \oint d(\arg G(\omega)) = 
   2 \pi \left( 
                 - \sum_{k=1}^{N_z} n_{z,k} + \sum_{k=1}^{N_p} n_{p,k} 
          \right) \;,
\end{equation}
where $N_z$ is the number of encircled zeros of $G(\omega)$, $N_p$ the number
of poles and $n_{z,k}$ and $n_{p,k}$ their multiplicities.  For example, the
function $G(\omega)=i\omega \gamma$ has a single zero ($N_z=1$) with
multiplicity 1 ($n_{z,1}=1$) at $\omega=0$.  Accordingly, it changes its phase
by $-2\pi$ on a clockwise circle around the origin. The function
$G(\omega)=-m\omega^2$, on the other hand, has a zero with multiplicity
$n_{z,1}=2$ at $\omega=0$ and it changes its phase by $-4\pi$ on the same
circle.

\begin{figure}[htbp]
  \begin{center}
    \includegraphics[width=16cm]{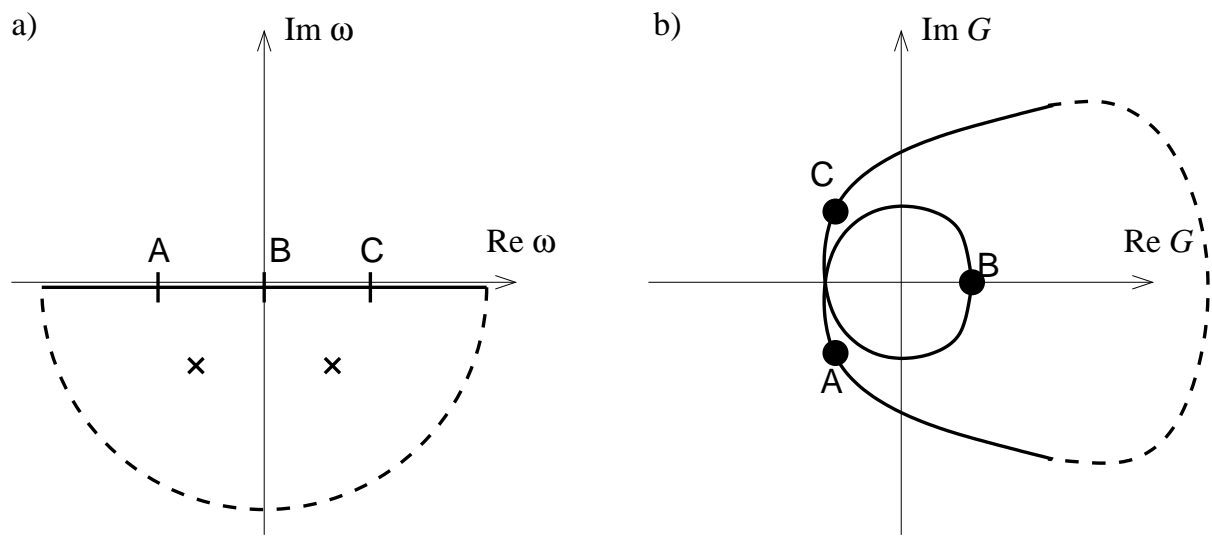}
  \end{center}
  \caption{a) The integration path for Eq.~\ref{eq:cauchy} in the complex
    $\omega$ plane. Encircled zeros (marked "$\times$") correspond to
    unstable eigenmodes. b) $G$ in the complex plane, with $\omega$ as
    a parameter.  The points denoted as A,B and C correspond to a
    negative, 0 and a positive value of $\omega$. The curve shown has
    a typical shape measured on insect flight muscle
    \cite{Machin.Pringle1960}.}
  \label{fig:chiplane}
\end{figure}

We apply Cauchy's theorem to a path that runs along the real $\omega$ axis from
$\omega=-\infty$ to $\omega=+\infty$ and close it in the negative imaginary
plane as shown in Fig.~\ref{fig:chiplane}a.  We further note that in a real
system $G(\omega)$ can have no poles in the negative imaginary half-plane,
because they would imply an infinite force response to a finite displacement
growing in time.  Therefore, any phase change along that path indicates a zero
with $\Im \omega <0$, and therefore a dynamical instability.

Now we can look at the same path in the complex $G$ plane, i.e., $\Im
G(\omega)$ plotted against $\Re G(\omega)$, where $\omega$ acts as a curve
parameter.  The strand along the real axis in the $\omega$ plane corresponds to
the $(\Re G(\omega), \Im G(\omega))$ curve, with $\omega$ taking real values
values from $-\infty$ to $\infty$.  In case $G(-\infty)\ne G(\infty)$, we close
the path in the positive real half-plane (Fig.~\ref{fig:chiplane}b), which
corresponds to the closure in the $\Im \omega<0$ half-plane.  A change of phase
of $G$ along the path in the $\omega$ plane by a multiple of $-2\pi$ directly
represents the same number of origin encirclements in the $G$ plane.
Therefore, it follows from (\ref{eq:cauchy}) that the complex $G(\omega)$ curve
encircles the coordinate origin $\sum_1^{N_z}n_{z,k}$ times in clockwise
direction.  

Note that the curve is symmetric with respect to the real axis.  This becomes
evident from the following consideration. Because $G(t)$ is a real function,
its Fourier transformed has to fulfill the following symmetry relation:
\begin{equation}
  \label{eq:symmetry}
G(-\omega^*)=G^*(\omega)\;.  
\end{equation}
For real $\omega$ values, this means $G^*(\omega)=G(-\omega)$, and for each $G$
value its complex conjugate is part of the curve as well.

An additional consequence of the Nyquist criterion is that the origin can only
be encircled in clockwise direction, because $G$ contains zeros but no poles in
the relevant region.

\section{Delayed force activation}

\begin{figure}[htbp]
  \begin{center}
    \includegraphics{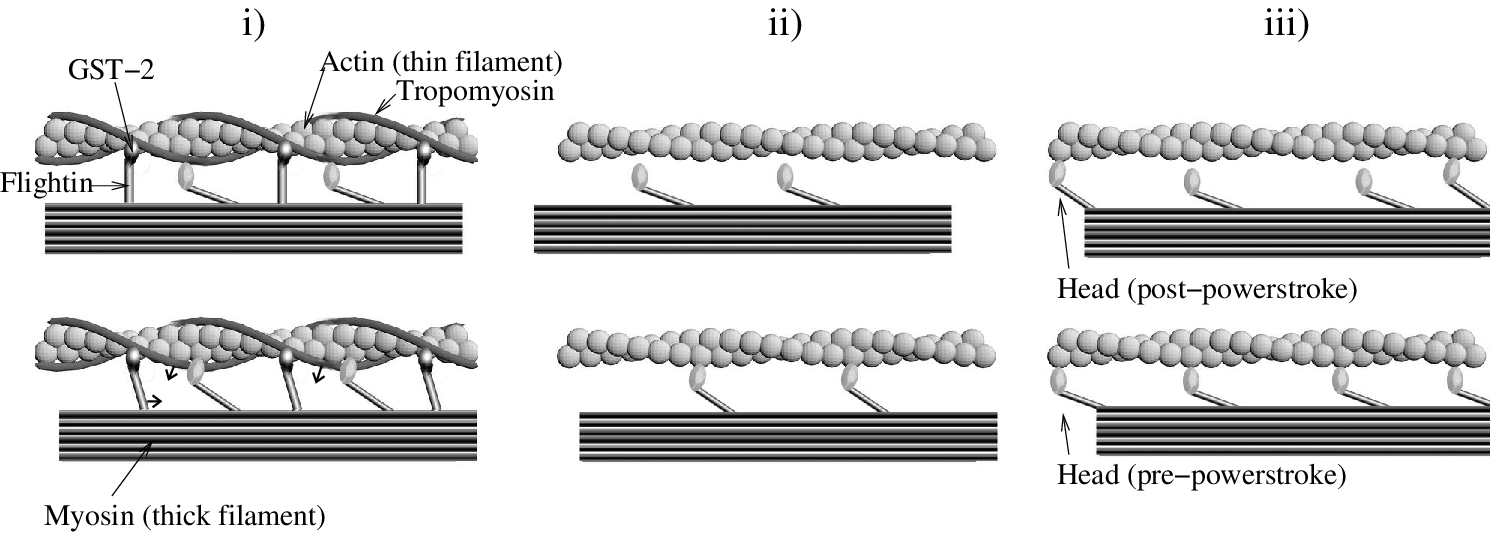}\\
    \includegraphics{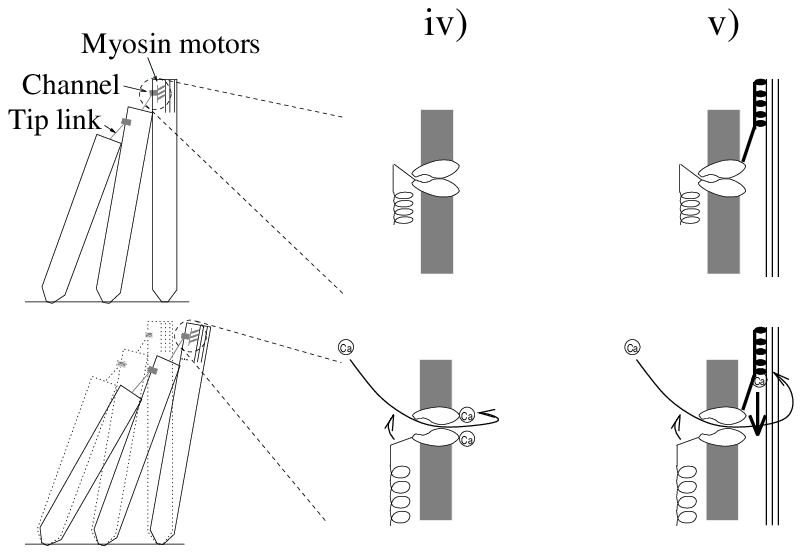}
  \end{center}
  \caption{Different biological mechanism that lead to delayed force activation:
    i) Stretch activation in insect flight muscle, after
    \cite{Maughan.Vigoreaux1999}. The applied stretch activates
    myosin mechanically through conformational changes involving
    flightin, GST-2, troponin-H and tropomyosin. ii) The alternative
    match-mismatch mechanism, where the myosin heads are activated
    when they are brought to appropriate positions along the actin
    helix. iii) The mechanism involving switching of myosin heads
    between two bound conformations.  A stretch brings more myosin
    heads into the pre-powerstroke state, which is less likely to
    detach from actin, and therefore the total number of active heads
    increases \cite{vilfan2003b}. iv) The fast adaptation mechanism in
    auditory hair cells.  When the transduction channel opens, calcium
    ions enter and promote the reclosure of the channel. v) The slow
    adaptation mechanism in auditory hair cells: calcium ions entering
    the cells reduce the force generated by myosin motors, therefore
    they slip back and the channel is more likely to close again.}
  \label{fig:mechanisms-delayed-feedback}
\end{figure}

One type of models, often used to explain oscillations generated by
molecular motors involves a delayed force activation mechanism.  This
means that when the system is displaced by the distance $x$ in one
direction, it develops a restoring force $f$, however not
instantaneously, but according to a differential equation like:
\begin{equation}
  \label{eq:delayedfeedback}
  \frac{df}{dt} = (K_A x -  f)/\tau \;.
\end{equation}
Here $\tau$ represents the time constant with which the force reacts to a
position change and $K_A$ an effective steady state stiffness of the active
system.  There are many different physical mechanisms that can produce a
delayed force response (Fig.~\ref{fig:mechanisms-delayed-feedback}):
\begin{enumerate}
\item Stretch-activation of insect flight muscle: a number of
  regulatory proteins facilitate the activation of myosin motors
  following a mechanical stretch (a relative displacement of thin
  filaments relative to thick)
  \cite{Pringle1978,Dickinson.Irving2005,Agianian.Bullard2004}.  After
  the activation, the development of force does not follow
  instantaneously, but with a certain time constant, typically with
  the attachment rate of motors.
\item In a sarcomere a position change can bring myosin heads closer
  to the accessible actin sites \cite{thomas98,Thomas.Thornhill1996}.
  This mechanism has been proposed for insect flight muscle
  \cite{Wray1979}, although experiments have later refuted it in many cases
  \cite{Squire1992}.
\item The switching of bound myosin heads between two conformations
  with different detachment rates can lead to an effective
  stretch-dependent activation
  \cite{vilfan2003b,Campbell.Slinker2001}.
\item The opening of a mechanically gated ion channel can cause the
  influx of calcium ions, which then promote channel reclosure.  This
  is called the fast adaptation mechanism in auditory hair cells and can
  generate spontaneous hair bundle oscillations
  \cite{Ricci.Fettiplace1998,Holt.Corey2000,Choe.Hudspeth1998,vilfan2003a}.
  If the hair cell is on the verge of spontaneous oscillations (i.e.,
  close to a Hopf bifurcation) it has the best sensitivity, dynamic
  range and frequency selectivity \cite{Camalet.Prost2000}. 
\item Calcium ions, entering the ion channel, can reduce the force
  generated by adaptation motors of type myosin 1C
  \cite{Batters.Molloy2004}, a mechanism known as slow adaptation in
  auditory hair cells
  \cite{Howard.Hudspeth1987,Martin.Hudspeth2000,Martin.Julicher2001}.
\item Some models for flagellar dynein are based on a
  strain-controlled \cite{Machin1958} or curvature-controlled
  \cite{Brokaw1971} activation.
\end{enumerate}

\begin{figure}[htbp]
  \begin{center}
    \includegraphics[width=16cm]{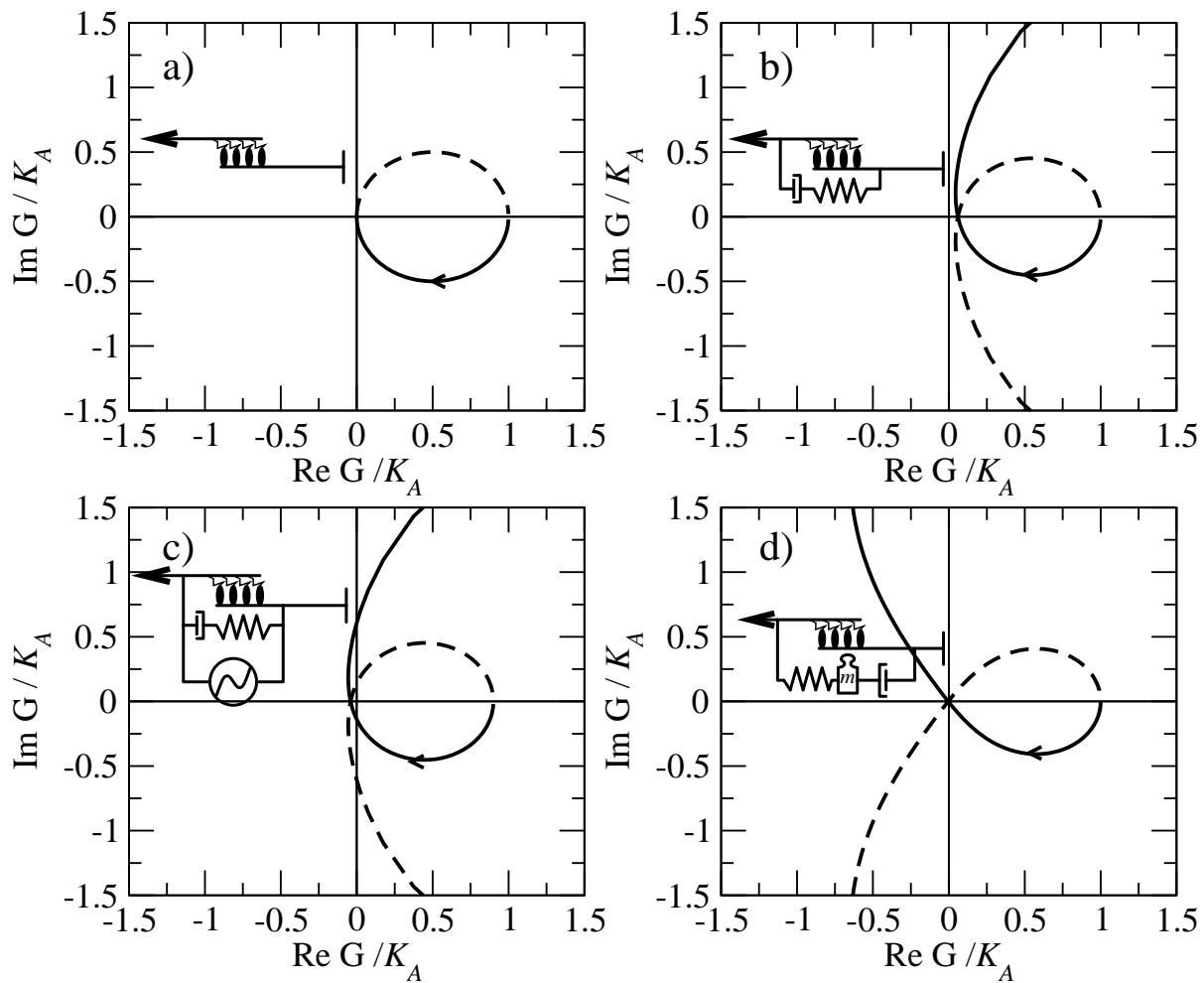}
  \end{center}
  \caption{Nyquist plots (also called Cole-Cole 
    plots or vector modulus plots) of the response function of a
    system with delayed force activation.  a) The response function of
    the delayed force activation alone, given by
    Eq.~(\ref{eq:chi_df}). b) Response function of the delayed force
    activation, combined with a damping element, $G_A+G_D$ (spring
    $K=5K_A$, dashpot with a constant $\gamma/K=100/\tau$).  c) System
    as in (b), additionally combined with an element with negative
    effective elasticity; $G_A+G_D+G_N$ ($K_1=-0.2 K_A$).  d) Response
    function $G_A+G_I$ of the system with delayed force activation,
    combined with a damped inertial oscillator ($\omega_0=10/\tau$,
    $K=2K_A$, $\beta=10/\tau$).}
  \label{fig:nyquist_sa}
\end{figure}

The solution of (\ref{eq:delayedfeedback}) can be expressed with a
Green's function $G_A(t)$, so that
\begin{equation}
  \label{eq:delayedfeedback-greeen}
  f(t)=\int_{-\infty}^{t} G_A(t-t') x(t') dt' \qquad \mbox{with} \qquad
  G_A(t)= \frac{K_A}{\tau} e^{-t/\tau} \;.
\end{equation}
Such behavior is well known from many branches in physics under the
term {\em Debye relaxator}. The corresponding response function
\begin{equation}
  \label{eq:chi_df}
  G_A(\omega)=\frac{K_A }{1 + i \omega \tau}
\end{equation}
is represented by a circle in a Nyquist plot (see
Fig.~\ref{fig:nyquist_sa} (a)).
It does have a negative imaginary part, however, the real part is always
positive. Hence the Nyquist plot of $G_A(\omega)$ cannot encircle the
coordinate origin and thereby satisfy the Nyquist criterion for an
instability.  The stationary state of the Debye relaxator simply is a
time-independent displacement $x^\star = f / K_A$. Note, that a dynamics with
a one-dimensional phase space, i.e.  first order differential equations of the
form $df/dt = x(f)$, is either monotonic or constant \cite{Strogatz1994}. Only
if the dynamic variable is on a circle or if the phase space has more than two
dimensions oscillations emerge as possible solutions.

One way to make the system unstable is to connect the active
mechanism with an element that shifts the Nyquist plot towards
negative values on the real axis; for an illustration of the type of
connection see Fig.~\ref{fig:nyquist_sa}. This can be trivially
achieved through an element with a negative effective stiffness, or,
as we will show below, through an inertial load.

Figure \ref{fig:nyquist_sa} shows the response function of the delayed
activation system (\ref{eq:chi_df}), combined with different elements.
The first is an element consisting of a spring (constant $K$) and a
dashpot (damping coefficient $\gamma$) in series.  Their combined
response function reads
\begin{equation}
  \label{eq:chi-spr-dash}
  G_D(\omega)=\left(\frac 1 K + \frac 1 {i\omega \gamma} \right)^{-1} \;.
\end{equation}
A second element we connect to the motor assembly is an element with
negative effective elasticity,
\begin{equation}
  \label{eq:chi-negst}
  G_N(\omega)=K_1 < 0\;.
\end{equation}
The third element is a damped inertial oscillator (a spring connected
to a mass, with an additional damping element connecting both to a
fixed point). The response function of such an element reads
\begin{equation}
  \label{eq:chi-inert}
  G_I(\omega) 
 = \left( \frac 1 K + \frac{1}{-m \omega^2 + i \omega \gamma} \right)^{-1}
=K\left( 1- \frac{\omega_0^2}{\omega_0^2-\omega^2+i \beta \omega} \right)
\end{equation}
with $\omega_0^2=K/m$ and $\beta=\gamma/m$.

Negative stiffness can be generated by the gating compliance of ion
channels \cite{Howard.Hudspeth1988,Martin.Hudspeth2000} or by motors
switching between two states \cite{hill74,vilfan2003b}.  Inertial load
is the likely mechanism of ensuring an instability in insect flight
muscle \cite{vilfan2003c}. Also in many types of hair cells, no
negative stiffness has been observed, but as the hair bundle is often
attached to the tectorial membrane, inertial load could play an
important role as well \cite{Authier.Manley1995}.

\section{Anomalous force-velocity relations}

A class of models that were frequently discussed in relation to
dynamical instabilities in systems of molecular motors is based on an
anomalous force-velocity relation, meaning that a group of coupled
motors produces a force that displays a region with negative slope as
a function of velocity.  Models for dynein \cite{Brokaw1975}, myosin
\cite{vilfan99}, propulsion by actin polymerization
\cite{Bernheim-Groswasser2005}, mitotic spindle oscillations
\cite{Grill.Julicher2005} as well as generic motors
\cite{prost95,juelicher97} were based on this mechanism.

A common feature of such models is translational invariance, meaning
that the force $f(t)$ does not depend on the position $x$, but only on
the velocity $v$ and the history thereof.  The translational
invariance can be achieved even on periodic tracks (like actin
filaments) if the motors (myosin heads) are arranged incommensurate to
the track periodicity \cite{prost95}.

An instructive example of a system displaying an anomalous force-velocity
relation is a two-state cross-bridge model for motors like myosin. In this
model, motors can attach to actin in a forward-leaning position with rate
$r_{\rm a}$, then quickly undergo a conformational change, which makes them
strained.  When the filaments moves, this strain changes with time (for
forward-running motors, it decreases), and when the motor eventually detaches
with a strain-dependent rate $r_{\rm d}(\xi)$ the cycle can repeat.  Each motor
generates a force $k\xi$, proportional to its strain $\xi$. A full description
of this system requires Master equations for the probability of a motor for
being in the detached state ($\Phi_{\rm d}$) and for the probability density of
a motor for being in the attached state with strain $\xi$ ($\Phi_{\rm
  a}(\xi)$).  \numparts
\begin{eqnarray}
\label{eq:master}
\left(\partial_t- v  \partial_\xi\right)\Phi_{\rm a}(\xi,t) &=&
\Phi_{\rm d}(t) r_{\rm a} P(\xi)
-\Phi_{\rm a}(\xi,t) r_{\rm d}(\xi) \\
\partial_t\Phi_{\rm d}(t) &=& -\Phi_{\rm d}(t) r_{\rm a} +\int_{-\infty}^\infty d\xi \,
 \Phi_{\rm a}(\xi,t) r_{\rm d}(\xi) \, ,
\end{eqnarray}
\endnumparts
Both probabilities are not independent but normalized such that
$\Phi_{\rm d}+\int_{-\infty}^\infty \Phi_{\rm a}(\xi) d\xi=N$ gives the total number of
motors.  $P(\xi)$, normalized as $\int_{-\infty}^\infty P(\xi)d\xi=1$ is the
distribution of strains on newly attached motors, with an expectation value
$d=\int_{-\infty}^\infty P(\xi) \xi d\xi$.  A general analytical solution for
the stationary state of these equations is given in \cite{vilfan99}.

\begin{figure}[htbp]
  \begin{center}
    \includegraphics[width=16cm]{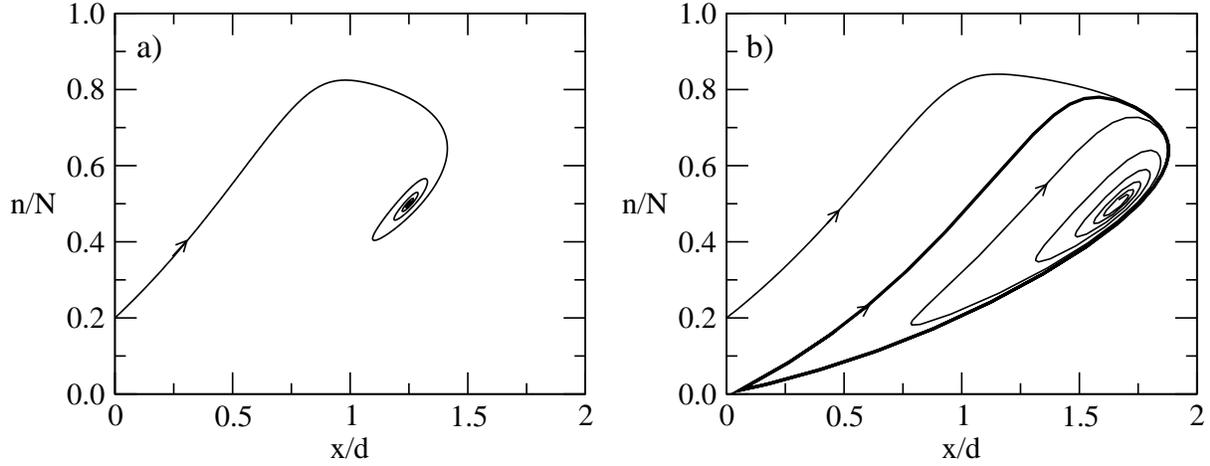}
  \end{center}
  \caption{Numerical solution of the equations of motion
    (\ref{eq:eq_motion},b) for the system of motors, coupled to a spring
    with a constant $K=0.4 \times N k$ (a) and $K=0.3\times Nk$ (b).
    The force-dependent detachment rate was chosen as $r_d(y)=r_a \exp(2
    (x/d)^2-2)$.  The system has a stable fixed point in the first case and an
    unstable one, surrounded by a limit cycle, in the second case.}
  \label{fig:trajectories}
\end{figure}

To illustrate the key physical features of this model we neglect the
strain-dependence of the attachment probability and assume that all
attached motors have the same strain $y(t)$,
\begin{eqnarray}
  \Phi_{\rm d} (\xi, t) = n (t) \, \delta (\xi - y(t)) \;.
\end{eqnarray}
Then the number of attached motors $n(t)$ is related to the number of detached
motors by $\Phi_{\rm d} (t) = N - n(t)$. Upon inserting these relations into
Eqs.\ref{eq:master},b , multiplying Eq.\ref{eq:master}
with $\xi$ and integrating over all values of $\xi$ one obtains a simplified
set of dynamic equations for $y(t)$ and $n (t)$: \numparts
\begin{eqnarray}
  \label{eq:eq_motion}
  \dot y &=& [d-y] \frac{N-n}{n} r_{\rm a} -v\\
  \dot n &=& [N-n] r_{\rm a}  -n r_{\rm d}(y)
\end{eqnarray}
\endnumparts
The external force on the system, which equals the total force produced by the
motors, is given as
\begin{equation}
  \label{eq:force1}
  f=- n k y
\end{equation}
where $k$ denotes the spring constant of each motor.  These are nonlinear flow
equations with a two-dimensional phase space consisting of the strain $y(t)$
and the number density $n(t)$ of the attached motors. For a constant velocity
$v$, they always have a stable stationary solution.  However, if we take the
system consisting of the group of motors, coupled to an external spring with a
spring constant $K$, so that the force equilibrium states $f+Kx=0$, we obtain a
new system of two nonlinear equations for two independent variables (e.g., $x$
and $n$).  Depending on the parameter values, they can have a stable fixed
point or a limit cycle (oscillatory steady state).  These two examples are
shown in Fig.~\ref{fig:trajectories}.

In the following we will again use the stability analysis of linear response
functions to characterize the system.  In the absence of an external velocity
($v=0$), Eqs.~\ref{eq:eq_motion},b have a stable stationary solution with
$y_0=d$ and $n_0=\frac{r_{\rm a}}{r_{\rm a}+r_{\rm d}(d)}N$.  For small values
of $v$ we can linearize the flow equations to \numparts
\begin{eqnarray}
  \label{eq:motion_linearized}
  \dot y &=& -r_{\rm d}(d)\Delta y - v\\
  \dot n &=& -n_0 r_{\rm d}'(d) \Delta y - (r_{\rm a}+r_{\rm d}(d)) \Delta n
\end{eqnarray}
\endnumparts where $\Delta y = y -y_0$ and $\Delta n = n -n_0$. These
equations have two negative eigenvalues, $-\lambda_1 =-r_{\rm d}(d)$ and
$-\lambda_2=-r_{\rm a}-r_{\rm d}(d)$.  The solution of the coupled system then
has the form \numparts
\begin{eqnarray}
  \label{eq:solution_motion}
   \Delta y(t)  &=&
   - \int_{-\infty}^{t}  dt' e^{-\lambda_1 (t-t')} v (t') \\
   \Delta n(t) &=&  \frac{n_0 r_{\rm d}'(d)}{r_{\rm a}} \, 
   \int_{-\infty}^{t}  dt' \left[ e^{-\lambda_1 (t-t')} v (t')
                                       - e^{-\lambda_2 (t-t')} \right] v (t')
\end{eqnarray}
\endnumparts
The force can be determined as
\begin{equation}
  \label{eq:force_solution}
  f(t) = -nky
  = f_0 + \int_{-\infty}^{t} dt' k n_0 
    \left[ (1-\alpha) e^{-\lambda_1 (t-t')} + \alpha
            e^{-\lambda_2 (t-t')}  
    \right] v (t')  \;.
\end{equation}
Where $\alpha=\frac{d r_{\rm d}'(d)}{r_{\rm a}}$ is a dimensionless coefficient.
With a more precise solution, or with more complex model equations,
the solution could contain more than two relaxation constants, but its
basic form would remain.  Inserting a stationary velocity ($v(t)\equiv
\mbox{const}$) in Eq.~(\ref{eq:force_solution}) one can see that the
system displays an anomalous stationary force velocity relation if
$\alpha$ is sufficiently large that
\begin{equation}
  \label{eq:criterion}
  \frac{1-\alpha}{\lambda_1}+\frac{\alpha}{\lambda_2}<0
\end{equation}

\begin{figure}[htbp]
  \begin{center}
    \includegraphics[width=16cm]{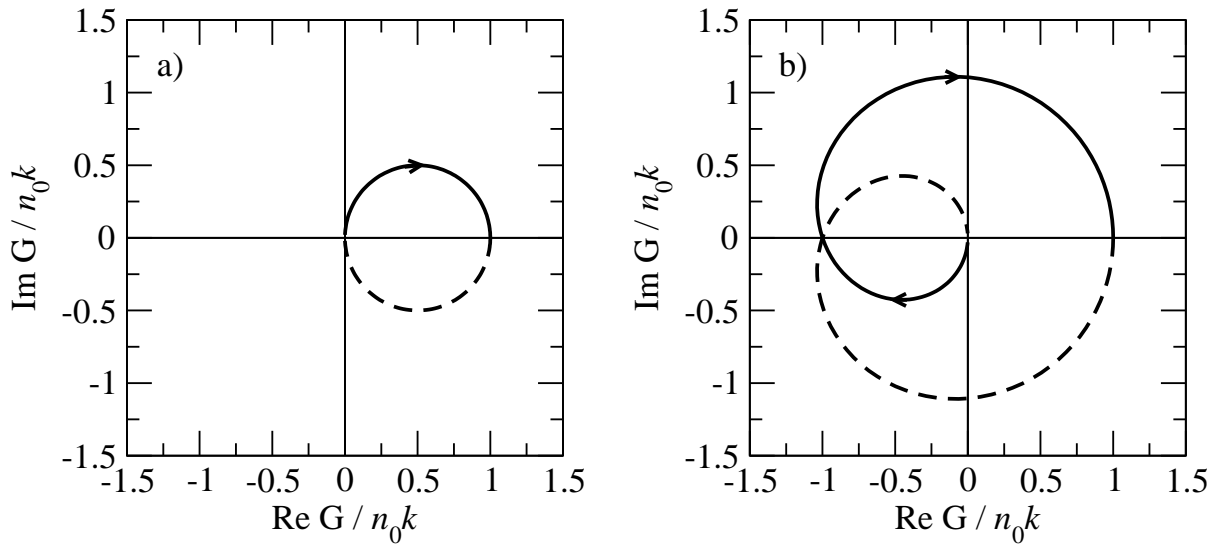}
  \end{center}
  \caption{Nyquist plots (Cole-Cole plots, vector modulus plots) 
    of the response function $G(\omega)$ (\ref{eq:chi_fv}) for
    $\alpha=0$ (a) and $\alpha=5$ (b).  Other parameter values:
    $r_{\rm d}(d)=r_{\rm a}$, implying $\lambda_2=2\lambda_1$.  Negative $\omega$
    values are shown with dashed lines. The system in (a) is
    dynamically stable, because the curve does not encircle the
    origin, while that in (b) oscillates spontaneously.}
  \label{fig:fv-nyquist}
\end{figure}

The response function in Fourier space follows from
Eq.~(\ref{eq:force_solution}):
\begin{equation}
  \label{eq:chi_fv}
  G(\omega) = 
  n_0 k (1-\alpha) \frac{i\omega}{\lambda_1+i\omega} +n_0 k \alpha
  \frac{i\omega}{\lambda_2+i\omega} 
\end{equation}
In the limit $\omega \to \infty$, the response is always that of the
elastic elements involved, $G=n_0 k$.  Figure \ref{fig:fv-nyquist}
shows the resulting Nyquist plot for two scenarios: with a normal and
an anomalous force-velocity relation.  In case of an anomalous force
velocity relation the curve encircles the origin and the system is
dynamically unstable.

If the assembly of motors is coupled to an elastic element (spring constant
$K$), the curve in the Nyquist plot is shifted to the right by the amount $K$.
There is a critical value of $K$ at which the oscillations stop.  In the
example shown in Fig.~\ref{fig:trajectories}, this value would lie between
those used in both diagrams. At the transition point the system exhibits a
Hopf bifurcation \cite{juelicher97,Grill.Julicher2005}. 

In summary, based on the foregoing discussion, we feel that a
combination of methods from the engineering sciences like control
theory and methods from the theory of dynamic systems may be fruitful
for future analysis of molecular motor assemblies or other functional
units in cellular systems.

\subsection*{Acknowledgment}
We have benefitted from discussions with Tom Duke, Frank J\"ulicher,
Thomas Franosch and Franz Schwabl.  This work was supported by the
Slovenian Office of Science (Grants No.~Z1-4509-0106-02 and
P0-0524-0106).

\end{document}